# Probing Dynamics of Majorana Fermions in Quantum Impurity Systems

Christophe Mora[1] and Karyn Le Hur[2]

[1] *Laboratoire Pierre Aigrain, École Normale Supérieure,*
*Université Paris 7 Diderot, CNRS; 24 rue Lhomond, 75005 Paris, France and*
[2] *Center for Theoretical Physics, Ecole Polytechnique, CNRS, 91128 Palaiseau Cedex, France*

We investigate the admittance of a metallic quantum RC circuit with a spinful single-channel lead or equally with two conducting spin-polarized channels, in which Majorana fermions play a crucial role in the charge dynamics. We address how the two-channel Kondo physics and its emergent Majoranas arise. The existence of a single unscreened Majorana mode results in non-Fermi liquid features and we determine the universal crossover function describing the Fermi-liquid to non-Fermi liquid region. Remarkably, the same universal form emerges both at weak transmission and large transmission. We find that the charge relaxation resistance strongly increases in the non-Fermi liquid realm. Our findings can be measured using current technology assuming a large cavity.

The need for fast manipulation and readout of quantum coherent circuits, notably in the perspective of quantum computation, has been a strong motivation to investigate the dynamical response of nanoconductors [1]. Excited at frequencies $\omega$ in the quantum regime, $\hbar\omega \gg k_B T$, the systems evolve due to the intriguing interplay of correlation and quantum coherence effects. The quantum RC circuit [2], a quantum dot attached to a single lead and polarized by an external AC gate voltage, has emerged as the archetypical system for studying the dynamics of coherent circuits [3–11]. Recent experiments [12] on quantum hybrid structures combining microwave resonators [13] with semiconductor or nanotube quantum dots offer an alternative perspective to measure the admittance of quantum circuits [14]. The existence of a quantized [15] charge relaxation (AC) resistance $R_q = h/(2e^2)$ in the quantum RC circuit has been shown [16] (and measured [2]) to originate from the Fermi liquid (FL) nature of low energy excitations where the elementary quasiparticles are non-interacting fermions. A deep connection [3] has also been drawn between the quantized resistance $R_q = h/e^2$ for large dots, the Shiba relation and the one-channel Kondo model in RC circuits [17] close to the charge degeneracy points.

In this paper, we investigate the non-Fermi liquid (NFL) situation where the elementary quasiparticles are Majorana fermions [18]. This description naturally applies to the quantum RC circuit with spinful (spin unpolarized) electrons and a large cavity (dot) [19–21] and is associated to the two-channel Kondo model [22]. It could be extended to the case of the helical edges of quantum spin Hall states [23–25] since the model is invariant upon reversing the direction of one of the spin species [26]. As discussed below, the corresponding low energy effective theory involves eight chiral Majorana fermions [27] and a local Majorana fermion (Klein factor) representing the residual spin of the impurity. Although the local Majorana can not be manipulated as a separate object and used for quantum computation, its presence is fundamental in the emergence of NFL physics [28].

The search for the existence of Majorana fermions has engendered a spurt of experimental efforts in condensed-matter systems [29–35]. In our case, the local Majorana is a remnant spin degree of freedom and not a composite object resulting from superconductivity as in topological wires [36]. Nevertheless, our system is described at low energy by a Majorana resonant level model, or the Emery-Kivelson model [37], which also describes the coupling of a local Majorana fermion to a normal lead in a topological superconducting wire.

In the quantum RC circuit with two conducting (spin) modes, the local Majorana fermion acquires a spectral width $\Gamma$, due to its coupling to the leads, which sets a crossover energy scale. Below $\Gamma$, the dynamics of the local Majorana is quenched and FL physics dominates while NFL behaviour [38] emerges at energies above $\Gamma$. The crossover energy scale $\Gamma$ vanishes at the charge degeneracy points [39]. Here, we provide an analytical expression for charge fluctuations along this universal crossover as a function of frequency: the charge relaxation resistance starts at $R_q = h/(2e^2)$ for a '2-mode' large cavity when $\omega = 0$ and rapidly increases with frequency towards the NFL region.

The system under study comprises a large (metallic) quantum dot attached to a lead via a quantum point contact (QPC) with a single spin-unpolarized channel [19, 20, 39, 40]. The quantum RC circuit could be equally built at the helical edges of quantum spin-Hall insulators [11]. Electron confinement implies a charging energy $E_C = e^2/(2C_g)$, where $C_g$ is the capacitance of the dot, and the interaction term $H_C = E_C(\hat{N} - N_0)^2$ in the Hamiltonian. $N_0$ is the dimensionless gate voltage and the operator $e\hat{N}$ gives the electron charge on the dot. Below, we address the extreme cases of almost transparent and weakly transmitting QPC.

We consider first an almost open dot with weak charge quantization, *i.e.* charge quantization is strongly smeared out by the large dot-lead coupling. The model can be reduced to a one-dimensional form with coordinate $x$, the region $x < 0$ defining the lead and $x > 0$ the (infinite) dot. Electrons are weakly backscattered, with amplitude $r \ll 1$, at the boundary $x = 0$. In this regime, spin and charge excitations occur at well-separated energy scales, $r^2 E_C \ll E_C$, and the system is conveniently described using bosonization [41, 42] in the spin and charge sectors. Following a standard sequence [39, 43] of bosonization and refermionization (see Supplementary materials), we find the exact action describing the system $S = S_F + S_c + S_{BS}$, with

$$S_c = \sum_{\omega_m} |\phi_c(\omega_m)|^2 \left( |\omega_m| + \frac{2E_C}{\pi} \right), \tag{1a}$$

$$S_{BS} = ir_0 \int_0^\beta d\tau\, \eta(\tau)\, \hat{a}(\tau) \cos\left( \sqrt{2}\phi_c(\tau) + \pi N_0 \right), \tag{1b}$$

where $r_0 = 2v_F\, r\, \sqrt{2/\pi a_0}$, $\eta(\tau) \equiv \eta(x = 0, \tau)$ and $\phi_c(\tau) \equiv \phi_c(x = 0, \tau)$. The charge bosonic field at $x = 0$ is related to the charge on the dot $\phi_c = (\pi/\sqrt{2})\hat{N}$. The bosonization procedure introduces a boson field $\phi(x)$ which embodies spin excitations along the one-dimensional fermionic line. Refermionization of $\phi$

$$\frac{1}{2\sqrt{\pi a_0}} e^{i\phi(x)} = \hat{a}\, \psi(x), \tag{2}$$

defines a (Klein factor) Majorana fermion $\hat{a} = \hat{a}^\dagger$. In Eq. (1a), $S_F$ is the free part for the chiral Majorana fermion $\eta(x) = (\psi(x) - \psi^\dagger(x))/(i\sqrt{2})$. Here the Majorana $\hat{a}$ has nothing to do with superconductivity but rather describes the residual spin-1/2 degree of freedom emerging after the Kondo screening of the original spin-1/2 at the dot-lead interface [39, 43]. The existence of this unscreened degree of freedom is responsible, as discussed below, for the emergence of NFL features.

Integrating the massive charge field $\phi_c$ in Eq. (1a) yields, to leading order in $r \ll 1$, an exactly solvable Majorana resonant level model [39]. The Majorana fermion $\hat{a}$ acquires the spectral width $\Gamma = (8E_C\gamma/\pi^2)\, r^2 \cos^2(\pi N_0)$ with $\Gamma \sim r^2 E_C \ll E_C$ and $\ln\gamma = \mathbf{C} \simeq 0.5772$ is the Euler constant. Below the energy scale $\Gamma$, spin excitations are quenched. NFL features arise due to the combined effect of spin excitation, described by $\hat{a}$, with the quenching of charge excitation, that is for energies between $\Gamma$ and $E_c$. Below $\Gamma$, a crossover to a Fermi liquid regime is established [39].

We are interested in the charge susceptibility $\chi_C(t) = i\theta(t)\langle[\hat{N}(t), \hat{N}]\rangle$. The low frequency expansion of the related admittance

$$G(\omega) = -i\omega e^2 \chi_C(\omega) \equiv -i\omega C_0 \left(1 + i\omega C_0 R_q\right) \tag{3}$$

defines the differential capacitance $C_0$ and the charge relaxation resistance $R_q$. In order to compute $\chi_C$, the charge field $\phi_c$ should not be fully integrated and we need to extend the analysis of Refs. [39, 44]: this is discussed in the SM. Following a lengthy but straightforward perturbative calculation, for $r \ll 1$, we obtain at zero temperature [45] the result $\chi_C = K_0 + K_1 + K_2$,

$$e^2 K_0(\omega)/C_g = \alpha(\omega), \tag{4a}$$

$$e^2 K_1(\omega)/C_g = -8\gamma v_F\, r^2 \sin^2(\pi N_0)\, \alpha(\omega)^2\, \Pi_{a\eta}(\omega), \tag{4b}$$

$$e^2 K_2(\omega)/C_g = -8\gamma v_F\, r^2 \cos^2(\pi N_0)\, \alpha(\omega)^2\, \frac{\ln(E_C/\Gamma)}{2\pi} \tag{4c}$$

with $\alpha(\omega) = (1 - i\omega\pi/2E_C)^{-1}$. Only $K_0$ survives in the absence of backscattering $r = 0$, in which case, one obtains, by comparing with Eq. (3), $C_0 = C_g$ and $R_q = h/(2e^2)$, half of the result of spinless electrons for a large dot $h/e^2$ [3]. $\Pi_{a\eta}(\tau) = \langle\eta(\tau)\hat{a}(\tau)\eta\hat{a}\rangle$ is a polarization operator computed from the quadratic part of the action with the result

$$\Pi_{a\eta}(\omega) = -(1/2\pi v_F)\left[\ln(E_C/\Gamma) + \beta(\omega/\Gamma)\right],$$
$$\beta(x) = -(1 + 2i/x)\ln(1 - ix). \tag{5}$$

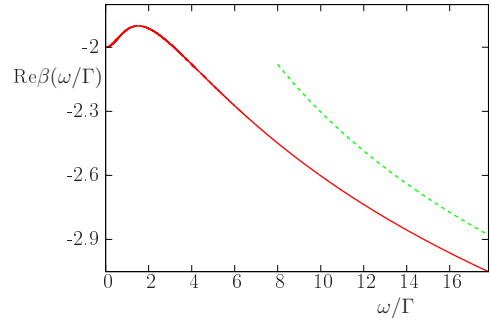

FIG. 1. Real part of the function $\beta$ as a function of the ratio $\omega/\Gamma$. The dotted line, $-\ln x$, gives the NFL asymptotical behaviour caused by the local Majorana fermion.

NFL behaviour in the charge susceptibility is signaled by logarithmic singularities in the computation of $\Pi_{a\eta}$ cutoff by the charging energy $E_C$. They arise in the contraction $\langle\hat{a}\hat{a}\rangle\langle\eta\eta\rangle$ and essentially originate from the fact that the Majorana operator $\hat{a}$ has zero dimension for energies between $\Gamma$ and $E_C$.

The function $\beta(\omega/\Gamma)$ describes the crossover between the FL and NFL responses for $\omega \ll \Gamma$ and $\omega \gg \Gamma$ respectively, its real part is shown in Fig. 1. For $\omega \ll \Gamma$, we use the expansion

$$\beta(x) \simeq -2 + x^2/6 + ix^3/6 - 3x^4/20 + \dots, \quad x \ll 1 \tag{6}$$

inserted in Eq. (4) and compare with Eq. (3) to extract $C_0$ and $R_q$. At vanishing frequency, the static susceptibility, and $C_0$, coincide precisely with Ref. [39]. Remarkably and similarly to the spinless case, we find no correction to the charge relaxation resistance $R_q = h/(2e^2)$ for $r \neq 0$ due to the absence of a linear term in Eq. (6). This result confirms FL behaviour at low energy. Indeed, using the Fermi liquid approach elaborated in Ref. [16], where lead electrons are coherently backscattered with a phase shift proportional to the static charge susceptibility [41], one easily derives the Shiba relation and show that $R_q = h/(2e^2)$ for an arbitrary transmission of the QPC. Note that $\Gamma$ vanishes at $N_0 = 1/2$ where the system is always a NFL.

We now turn to the opposite limit of weak transmission of the QPC. The system is adequately described [3, 17] by the tunnel Hamiltonian $H = H_0 + H_C + H_T$ where

$$H_T = t \sum_{k,k',s=\uparrow,\downarrow} \left( d^\dagger_{ks} c_{k's} + c^\dagger_{k's} d_{ks} \right) \qquad (7)$$

transfers electrons between the (large) dot and the lead with operators $c_k$ and $d_k$ respectively; the index $s$ refers, e.g., to the two spin polarizations. The free electron part reads $H_0 = \sum_{a=c,d,s} \varepsilon_k a^\dagger_{ks} a_{ks}$ for dot and lead electrons. $H_T$ either decreases or increases the dot charge by one unit and thus does not commute with $H_C$. Far from charge degeneracy, the perturbative approach of Ref. [3] can be reproduced with an additional factor two that accounts for spin degeneracy. One readily obtains $R_q = h/(2e^2)$, again in agreement with the Fermi liquid picture. Perturbation theory however breaks down close to charge degeneracy $N_0 \simeq 1/2$ where NFL physics starts to play a role. In this region, the charge states other than 0 and 1 can be disregarded and a mapping to the two-channel Kondo model formulated [3, 17] where the two charge states are represented by a spin 1/2 with $\hat{N} = \frac{1}{2} + S_z$. The vicinity to charge degeneracy $h_0 = E_C(1 - 2N_0)$ defines a local magnetic field coupled to $S_z$. Our study of charge fluctuations is then translated to a study of the local spin susceptibility in the two-channel Kondo model.

For $h_0 \ll T_K$, where $T_K$ is the Kondo temperature, two regimes have been identified [22] in the renormalization group (RG) analysis: NFL properties dominate for frequencies (energies) $\Gamma = h_0^2/(2T_K) < \omega < T_K$ while a FL response is obtained at smaller frequencies $\omega < \Gamma$. The crossover is investigated analytically using the SO(8) representation [27] of the two-channel Kondo model, which provides a simple description of the NFL fixed point [46, 47]. The bulk fermions, with two spin species and two channels, have a non-local representation in terms of eight chiral Majorana fermions. With no impurity, the free Hamiltonian reads

$$H_0^K = \frac{-iv_F}{2} \sum_{j=1}^{8} \int_{-\infty}^{+\infty} dx \, \chi_j(x) \partial_x \chi_j(x). \qquad (8)$$

The Majorana fermions $\chi_{1,2,3}$ generate the spin current, $\chi_{4,5,6}$ the flavor current and $\chi_7$, $\chi_8$ the charge current.

In the presence of the Kondo impurity coupled only to the spin current, the NFL infrared fixed point is simply characterized by the twisted boundary conditions $\chi_j(0^-) = -\chi_j(0^+)$ for $j = 1, 2, 3$. Absorbing this $\pi/2$ phase shift into a redefinition of the fields, $\chi_{1,2,3}(x) \to \text{sgn}(x)\chi_{1,2,3}(x)$, one recovers the free Hamiltonian Eq. (8) also at the infrared fixed point. A finite local magnetic field $h_0$ destabilizes this fixed point with the relevant perturbation [46] $H_b = i(h_0/\sqrt{T_K/v_F})\chi_1\hat{a}$ of scaling dimension 1/2, where $\chi_1 = \chi_1(0)$. The local Majorana fermion $\hat{a}$ describes the residual impurity spin. The Hamiltonian $H_{IR} = H_0^K + H_b$ is equivalent to the two-dimensional Ising model with a boundary magnetic field, a correspondence that has been used to calculate the one-body Green's function along the FL to NFL crossover [47]. For energies $\omega \ll \Gamma$, the relevant boundary term $H_b$ restores the continuity of $\chi_1(x)$ at $x = 0$. We thus recover a Fermi liquid as the even number of twisted fields ($\chi_2$ and $\chi_3$) indicates [27].

Quite generally, the impurity spin can be expanded over the different operators allowed by conformal field theory (CFT). At low energy, the leading term is

$$S_z = i\sqrt{\frac{v_F}{T_K}} \, \chi_1\hat{a}, \qquad (9)$$

in accordance with $H_b$. The spin susceptibility $\chi_s(\tau) = -(v_F/T_K)\langle\chi_1(\tau)\hat{a}(\tau)\chi_1\hat{a}\rangle$ is obtained by noting the equivalence between the Hamiltonian $H_{IR}$ and the quadratic action $S_F + S_{BS}^0$, derived in the large transparency case. We identify $\eta = \chi_1$ and $2\Gamma = h_0^2/T_K$ and find

$$\chi_{s0}(\omega) = \frac{1}{2\pi T_K}\left[\ln\left(\frac{T_K}{\Gamma}\right) + \beta(\omega/\Gamma)\right], \qquad (10)$$

describing the FL-NFL crossover. At large frequency $\omega \gg \Gamma$, the spin susceptibility exhibits NFL features

$$\chi_{s0}(\omega) = \frac{1}{2\pi T_K}\left[\ln\left(\frac{T_K}{\omega}\right) + \frac{i\pi}{2}\right], \qquad (11)$$

in agreement with the prediction of Conformal Field Theory [22] and Abelian bosonization [37]. The absence of a linear term in Eq. (6) requires, for the calculation of $R_q$, to include the leading irrelevant perturbation to $H_{IR}$,

$$\delta H = \frac{2\pi v_F^{3/2}}{\sqrt{T_K}} \, \chi_1\chi_2\chi_3\hat{a}. \qquad (12)$$

The only linear in frequency correction to the spin susceptibility comes from the vertex correction, shown Fig.2 a), $\delta\chi_s(\omega) = \chi_{s0}(\omega)\Pi_e(\omega)\chi_{s0}(\omega)$ where $\Pi_e(\omega)$ is the electron-hole pair susceptibility. At zero temperature, $\Pi_e(\omega) = i\pi\omega$, interpreted as the dissipation produced by electron-hole excitations. Expanding the spin susceptibility $\chi_s = \chi_{s0} + \delta\chi_s$ to linear order in $\omega$, we arrive at the Shiba relation

$$\text{Im}\chi_s(\omega) = \hbar\pi\omega \, \chi_s(0)^2, \qquad (13)$$

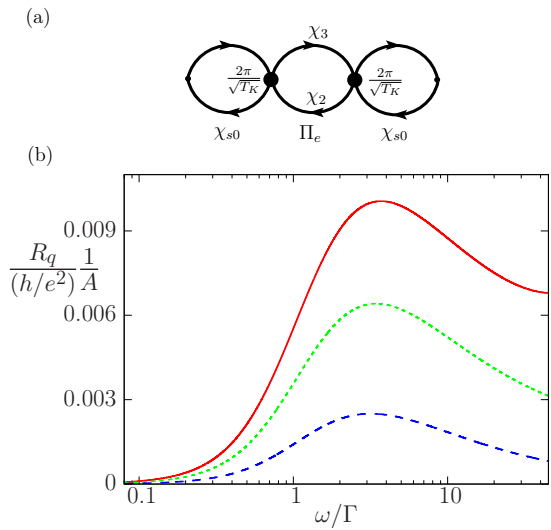

FIG. 2. (a) Vertex correction to the spin susceptibility $\chi_s$ to second order in $\delta H$. (b) Universal charge relaxation resistance valid at all transmissions showing the increase towards the NFL regime. It is plotted here for $B = 6, 7, 10$ (solid, dotted and dashed lines). At weak transmission, the maxima of $R_q$ are respectively 4.03, 7.02 and 55.1 in units of $h/e^2$. We note that the quantized result $R_q = h/2e^2$, recovered at zero frequency, is not visible in this plot computed in the scaling limit $A \gg 1$.

equivalent to the charge relaxation resistance $R_q = h/(2e^2)$. This result confirms the validity of the Fermi liquid picture [16] also at low transmission.

Finally, both for weak and almost perfect transparency, we examine the regime of intermediate frequencies $\omega \sim \Gamma$, where the expansion Eq. (3) is no longer relevant. Nevertheless, keeping $\omega \ll T_K, E_C$ and splitting the charge susceptibility into real and imaginary parts,

$$e^2 \chi_C(\omega) = C_0(\omega) + i\omega\, C_0(\omega)^2\, R_q(\omega) \qquad (14)$$

one can define frequency-dependent capacitance and charge relaxation resistance. This definition is relevant for experiments where the real and imaginary parts are measured separately [2]. At weak transmission and $\omega \ll T_K$, we extract the universal form

$$\frac{R_q(\omega)}{h/e^2} = A\, \Phi\left(\frac{\omega}{\Gamma}\right) = A\, \frac{\Gamma}{\omega} \frac{\mathrm{Im}\beta(\omega/\Gamma)}{[B + \mathrm{Re}\beta(\omega/\Gamma)]^2}, \qquad (15)$$

with $A = T_K/\Gamma$. The dimensionless function $\Phi(x)$ is plotted in Fig.2 b) for different values of $B = \ln(T_K/\Gamma)$.

Remarkably, the same scaling form involving the function $\Phi$ is obtained in the opposite limit of weak backscattering. In the scaling limit where $N_0 \rightarrow 1/2$ and $\omega \ll E_C$, one has $\alpha(\omega) \simeq 1$, $\sin(\pi N_0) \simeq 1$, $K_2 \simeq 0$ and $K_0 \ll K_1$ in Eqs. (4). As a result, one recovers Eq. (15) for the charge relaxation resistance $R_q$ with $B = \ln(E_C/\Gamma)$ and

$A = E_C/(4\gamma\, r^2\, \Gamma)$. The universality of the FL-NFL crossover has been anticipated by Matveev [39] who argued that the two-channel Kondo model influences the phase diagram beyond weak transparency [48].

To summarize briefly, we have shown that the presence of Majoranas in a quantum RC circuit results in a subtle charge dynamics which can, in principle, be revealed using current technology [45]. The FL to NFL crossover produces a visible increase of the charge relaxation resistance which can be probed through admittance measurements. We anticipate the possibility that NFL behaviour emerges also for a superconducting wire supporting Majorana fermions at his edges [29, 30, 32–36]. Majorana fermions can also be manipulated using a microwave cavity [49]. Other interesting directions concern the role of an asymmetry between channels which can be engineered through Zeeman effects for example [50].

We acknowledge discussions with I. Affleck, P. Dutt, M. Filippone, B. Horovitz, P. Le Doussal, Z. Ristivojevic, T. Schmidt and E. Sela. CM thanks L. Glazman for fruitful discussions and comments. KLH acknowledges support from DOE under the grant DE-FG02-08ER46541.

# Supplementary Information for
# Probing Dynamics of Majorana Fermions in Quantum Impurity Systems

Christophe Mora[1] and Karyn Le Hur[2]

[1] *Laboratoire Pierre Aigrain, École Normale Supérieure, Université Paris 7 Diderot,*
*CNRS; 24 rue Lhomond, 75005 Paris, France*
[2] *Center for Theoretical Physics, Ecole Polytechnique, CNRS, 91128 Palaiseau Cedex, France*

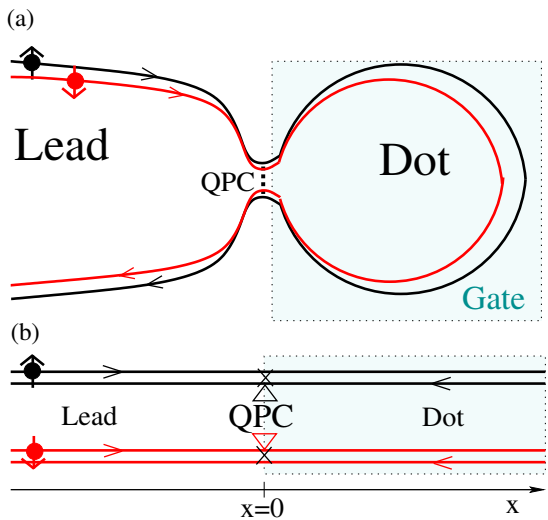

FIG. 1. (a) Schematic view of the two-channel quantum RC circuit considered in this article. Electrons move along two independent chiral edge states denoted here by the two spin orientations and represented in black and red colors. The quantum point contact separating the lead and dot regions is responsible for intra-channel backscattering at its position. The dot part of the circuit is coupled capacitively to a gate voltage resulting in a charging energy. (b) Reduced one-dimensional model describing the two-channel quantum RC circuit when the size of the dot is taken to infinity. In this limit, each chiral line gives independent right and left-moving states. The case of helical edge states for quantum spin Hall systems is easily obtained by reversing the direction of spin down electrons.

## I. DERIVATION OF THE EFFECTIVE ACTION

The RC circuit discussed in this work is a large quantum dot attached to a reservoir lead through a single-channel quantum point contact (QPC) [1], see Fig. 1. We concentrate on the spin-unpolarized case such that the two spin orientations give rise to two conduction channels for electrons, both in the lead and in the dot. The dot is large in the sense that its level spacing is the smallest energy scale in the problem [2, 3]. We emphasize that, in the case of an infinite dot, see Fig. 1(b), the direction of the spin down edge state can be reversed without altering the model discussed below. It then constitutes a natural description of helical edges of quantum spin Hall states [4].

The small size of the QPC, *i.e.* smaller than the Fermi wavelength, implies that it couples only the spherically symmetric (s-wave) parts of the wavefunctions in the lead and in the dot. Projecting on the s-wave component, the model can thus be given a one-dimensional form [2, 5, 6], with a coordinate $x$ such that the lead occupies the region $x < 0$ and the infinite dot the region $x > 0$. This choice is of course arbitrary. The Hamiltonien reads $H = H_0 + H_C + H_{BS}$,

$$H_0 = v_F \sum_\sigma \int_{-\infty}^{+\infty} dx \left[ \psi_{R\sigma}^\dagger(x)(-i\partial_x)\psi_{R\sigma}(x) \right. \tag{1a}$$

$$\left. + \psi_{L\sigma}^\dagger(x) i\partial_x \psi_{L\sigma}(x) \right] \tag{1b}$$

$$H_C = E_C \left( \hat{N} - N_0 \right)^2 \tag{1c}$$

$$H_{BS} = -v_F r \sum_\sigma \left( \psi_{R\sigma}^\dagger(0)\psi_{L\sigma}(0) + \text{h.c.} \right). \tag{1d}$$

$E_C = e^2/(2C_g)$ is the charging energy (see main text), $r$ the amplitude for electron backscattering at $x = 0$, $N_0$ the dimensionless gate voltage, and $v_F$ the Fermi velocity. The field operator $\psi_{R/L,\sigma}(x)$ describes a right(left)-moving fermion of spin $\sigma$ in the lead for $x < 0$ or in the dot for $x > 0$.

$$\hat{N} = \sum_\sigma \int_0^{+\infty} dx \left( \psi_{R\sigma}^\dagger(x)\psi_{R\sigma}(x) + \psi_{L\sigma}^\dagger(x)\psi_{L\sigma}(x) \right). \tag{2}$$

is integrated over all $x > 0$, it gives the number of electrons on the dot. This one-dimensional model can be bosonized (neglecting Klein factors for simplicity)

$$\psi_{R/L\sigma}(x) = \frac{1}{\sqrt{2\pi a_0}} e^{i[\pm\phi_\sigma(x) + \pi \int^x dy \Pi_\sigma(y)]}, \tag{3}$$

using the conjugated fields

$$[\phi_\sigma(x), \Pi_{\sigma'}(y)] = i\delta_{\sigma,\sigma'}\delta(x - y).$$

$a_0$ is a short-distance cutoff. For weak backscattering $r \ll 1$, the typical energy scales for charge and spin excitations, $E_C$ and $r^2 E_C$ respectively, are well-separated [2, 6]. This motivates the introduction of the charge and spin bosonic fields

$$\phi_c(x) = \frac{\phi_\uparrow(x) + \phi_\downarrow(x)}{\sqrt{2}}, \qquad \phi_s = \frac{\phi_\uparrow(x) - \phi_\downarrow(x)}{\sqrt{2}}, \tag{4}$$

and their respective conjugate momenta $\Pi_c(x)$ and $\Pi_s(x)$. With these new variables, one finds

$$
\begin{aligned}
H_0 =& \frac{v_F}{2\pi} \sum_{l=c/s} \int_{-\infty}^{+\infty} dx \left( \pi^2 [\Pi_l(x)]^2 + [\partial_x \phi_l(x)]^2 \right) \\
H_{BS} =& -\frac{2 v_F r}{\pi a_0} \cos\left( \sqrt{2} \phi_c(0) \right) \cos\left( \sqrt{2} \phi_s(0) \right),
\end{aligned}
\tag{5}
$$

and the charge of the dot is written only in terms of the charge field at $x = 0$, $\hat{N} = (\sqrt{2}/\pi)\, \phi_c(0)$.

The boundary term $H_{BS}$ involves $\phi_s(0)$ and thus couples only to the even part of the spin field, defined with its conjugated momentum as

$$
\begin{aligned}
\phi_1(x) &= \frac{\phi_s(x) + \phi_s(-x)}{\sqrt{2}}, \\
\Pi_1(x) &= \frac{\Pi_s(x) + \Pi_s(-x)}{\sqrt{2}},
\end{aligned}
\tag{6}
$$

on the semi-infinite positive axis $x \geq 0$. The problem is unfolded back on the complete $x$-axis by introducing the chiral field

$$
\phi(x) = \phi_1(|x|) - \pi \,\mathrm{sgn} x \int_0^{|x|} dy\, \Pi_1(y),
\tag{7}
$$

characterized by the canonical commutation relation $[\phi(x), \phi(y)] = i\pi \,\mathrm{sgn}(x - y)$. The Hamiltonian now reads

$$
\begin{aligned}
H_0 =& \frac{v_F}{2\pi} \int_{-\infty}^{+\infty} dx \left( \pi^2 [\Pi_c(x)]^2 + [\partial_x \phi_c(x)]^2 \right) \\
&+ \frac{v_F}{4\pi} \int_{-\infty}^{+\infty} dx \, [\partial_x \phi(x)]^2
\end{aligned}
\tag{8}
$$

where the odd part of the spin field has been omitted. The charging and tunneling terms take the form

$$
H_C = \frac{2 E_C}{\pi^2} \left( \phi_c(0) - \pi N_0/\sqrt{2} \right)^2
\tag{9a}
$$

$$
H_{BS} = -\frac{2 v_F r}{\pi a_0} \cos\left( \sqrt{2} \phi_c(0) \right) \cos \phi(0)
\tag{9b}
$$

In analogy with an impurity in a Luttinger liquid where the interaction parameter is $g = 1/2$ [7], the boundary operator $\cos\phi(0)$ has dimension 1 and can thus be refermionized with [5]

$$
\frac{1}{2\sqrt{\pi a_0}} e^{i\phi(x)} = \hat{a}\, \psi(x),
\tag{10}
$$

where the local Majorana fermion $\hat{a} = \hat{a}^\dagger$, and $\hat{a}^2 = 1/2$, has been introduced, which anticommutes with $\psi$. Note that $\hat{a}$ is a dynamical variable as it does not commute with the Hamiltonian. In term of the new fields $\psi$ and $\hat{a}$, we have

$$
\begin{aligned}
H_0 =& \frac{v_F}{2\pi} \int_{-\infty}^{+\infty} dx \left( \pi^2 [\Pi_c(x)]^2 + [\partial_x \phi_c(x)]^2 \right) \\
&+ \int_{-\infty}^{+\infty} dx \, \psi^\dagger(x)(-i v_F \partial_x)\psi(x)
\end{aligned}
\tag{11}
$$

and

$$
H_{BS} = \frac{2 v_F r}{\sqrt{\pi a_0}} \left( \psi(0) - \psi^\dagger(0) \right)\, \hat{a}\, \cos\left( \sqrt{2}\phi_c(0) \right)
\tag{12}
$$

We switch to an Euclidean action representation and integrate out the charge field $\phi_c(x)$ on the whole $x$-axis except at $x = 0$. The fermion spin field is decomposed in two chiral Majorana fermions $\psi(x) = \frac{\bar{\eta}(x) + i\eta(x)}{\sqrt{2}}$ and the second Majorana $\bar{\eta}(x)$ decouples completely from the charge modes. Omitting the field $\bar{\eta}$ and shifting $\phi_c$ to transfer the gate voltage $N_0$ to the tunneling term, the action finally reads $S = S_F + S_c + S_{BS}$ with

$$
\begin{aligned}
S_F =& \frac{1}{2} \int_0^\beta d\tau\, \hat{a}(\tau)\partial_\tau \hat{a}(\tau) \\
&+ \frac{1}{2} \int_0^\beta d\tau \int_{-\infty}^{+\infty} dx\, \eta(x, \tau)(\partial_\tau - i v_F \partial_x)\eta(x, \tau).
\end{aligned}
\tag{13}
$$

$S_c$ and $S_{BS}$ are given by Eqs. (1) in the main text, or Eq. (14a) below.

## II. CHARGE FLUCTUATIONS

### Derivation of the perturbative action

In order to compute the charge susceptibility, we need to extend the analysis of Refs. [2, 8]. Instead of integrating the action (Eq. (1) in the main text)

$$
S_c = \sum_{\omega_m} |\phi_c(\omega_m)|^2 \left( |\omega_m| + \frac{2 E_C}{\pi} \right),
\tag{14a}
$$

$$
S_{BS} = i r_0 \int_0^\beta d\tau\, \eta(\tau)\, \hat{a}(\tau)\, \cos\left( \sqrt{2}\phi_c(\tau) + \pi N_0 \right)
\tag{14b}
$$

over the whole charge mode $\phi_c$, we stop the integration at the intermediate energy scale $\Lambda$ such that $\Gamma \ll \Lambda \ll E_C$. By doing so, the massive part of the charge mode is almost fully integrated while we can still describe interactions between the spin and charge modes at energies $\sim \Gamma$. To leading order in $r \ll 1$, the factor $\cos\left( \sqrt{2}\phi_c(\tau) + \pi N_0 \right)$ in Eq. (14a) is substituted by its average

$$
\cos\left( \sqrt{2}\phi_c^l(\tau) + \pi N_0 \right)\, e^{-\langle \phi_c(\tau)\phi_c(\tau)\rangle_{\varepsilon > \Lambda}}
\tag{15}
$$

where $\langle \phi_c(\tau)\phi_c(\tau)\rangle_{\varepsilon > \Lambda} \simeq (1/2)\ln(\omega_D \pi/2 E_C)$, $\omega_D = v_F/\gamma a_0$ is the high energy cutoff of the initial model required by linearization of the spectrum [9]. The field $\phi_c^l(\tau)$ contains the low energy part of the charge mode.

The controlled expansion of the cosine in Eq. (15) is then justified by the reduction of available energies, notably

$$
\langle \phi_c^l(\tau)\phi_c^l(\tau)\rangle \sim (\Lambda/E_C)^2 \ll 1.
$$

A second order expansion yields the effective action $S = S_F + S_c^l + S_{BS}^0 + S_{BS}^1 + S_{BS}^2$ where the Coulomb interaction part $S_c^l$ keeps the structure of Eq. (14a) but with the frequency restriction $|\omega_m| < \Lambda$. Furthermore,

$$S_{BS}^0 = i\sqrt{2\Gamma v_F} \int_0^\beta d\tau\, \eta(\tau)\, \hat{a}(\tau),$$

$$S_{BS}^1 = -2i\sqrt{\Gamma v_F} \tan(\pi N_0) \int_0^\beta d\tau\, \phi_c^l(\tau)\, \eta(\tau)\, \hat{a}(\tau), \quad (16)$$

$$S_{BS}^2 = -i\sqrt{2\Gamma v_F} \int_0^\beta d\tau\, \left[\phi_c^l(\tau)\right]^2 \eta(\tau)\, \hat{a}(\tau).$$

In what follows, we consider the zero temperature limit $\beta \to \infty$. The action $S_F + S_c^l + S_{BS}^0$ defines an unperturbed quadratic problem, on top of which $S_{BS}^1$ and $S_{BS}^2$ act as perturbations. Using the time-ordered product $\mathcal{T}_\tau$, we introduce the Green's functions

$$
\begin{aligned}
G_{\hat{a}}(\tau) &= -\langle \mathcal{T}_\tau \hat{a}(\tau)\hat{a}(0)\rangle,\\
G_\eta(x,x',\tau) &= -\langle \mathcal{T}_\tau \eta(x,\tau)\eta(x,0)\rangle,\\
G_{\hat{a}\eta}(x,\tau) &= -\langle \mathcal{T}_\tau \hat{a}(\tau)\eta(x,0)\rangle,\\
G_\phi^l(\tau) &= \langle \mathcal{T}_\tau \phi_c^l(\tau)\phi_c^l(0)\rangle,
\end{aligned}
\quad (17)
$$

with the unperturbed expressions in Matsurbara space [3]

$$
\begin{aligned}
G_{\hat{a}}(i\omega_n) &= \frac{1}{i\omega_n + i\Gamma\mathrm{sgn}(\omega_n)},\\
G_\eta(0,0,i\omega_n) &= \frac{-i\mathrm{sgn}(\omega_n)}{2v_F}\frac{i\omega_n}{i\omega_n + i\Gamma\mathrm{sgn}(\omega_n)},\\
G_{\hat{a}\eta}(0,i\omega_n) &= -\sqrt{\frac{\Gamma}{2v_F}}\frac{\mathrm{sgn}(\omega_n)}{i\omega_n + i\Gamma\mathrm{sgn}(\omega_n)},\\
G_\phi^l(i\omega_n) &= \frac{\pi^2/(4E_C)}{1 + |\omega_n|\pi/(2E_C)}.
\end{aligned}
\quad (18)
$$

**Calculation of the polarization operator**

We provide below some details on the perturbative calculation of the charge susceptibility

$$\chi_C(\tau) = \frac{2}{\pi^2} G_\phi^l(\tau) = \frac{2}{\pi^2} \langle \mathcal{T}_\tau \phi_c^l(\tau)\phi_c^l(0)\rangle, \quad (19)$$

quoted in the main text, that is $\chi_C = K_0 + K_1 + K_2$. $K_0$ is the unperturbed result, obtained from the quadratic action $S_F + S_c^l + S_{BS}^0$. It does not depend on the backscattering amplitude $r$. $S_{BS}^1$ and $S_{BS}^2$ are included perturbatively and give $\chi_C$ a dependence on $r$. $K_1$ comes from the second order expansion of $S_{BS}^1$ and $K_2$ from the first order expansion of $S_{BS}^2$. We emphasize at this point that the small parameter of the expansion is not directly $r$ but the ratio $\Lambda/E_C \sim \Gamma/E_C$.

The result for $K_1$ is given at the second line of Eq. (4) in the main text, where the polarisation operator

$$\Pi_{a\eta}(\tau) = -\langle \mathcal{T}_\tau \hat{a}(\tau)\eta(\tau)\eta\hat{a}\rangle$$

has been introduced. We remind the notation $\eta(\tau) \equiv \eta(0,\tau)$. $\Pi_{a\eta}(\omega)$ is computed to leading order, that is from the unperturbed Green's functions Eq. (18). Two contractions are possible, where $\hat{a}$ is contracted with itself or with $\eta$. We denote

$$
\begin{aligned}
\Pi_1(\tau) &= G_{\hat{a}}(\tau)G_\eta(0,0,\tau),\\
\Pi_2(\tau) &= -G_{\hat{a}\eta}(0,\tau)G_{\hat{a}\eta}(0,-\tau),
\end{aligned}
\quad (20)
$$

the corresponding contributions. $\Pi_1$ is first computed for Matsubara frequencies ($\omega_m = 2\pi m T$)

$$\Pi_1(i\omega_m) = T \sum_n G_{\hat{a}}(i\omega_n)G_\eta(x=0, i\omega_n + i\omega_m). \quad (21)$$

Using standard complex-plane integration technics, the summation over the fermionic Matsubara frequencies, $\omega_n = \pi T(2n+1)$, is transformed into two contour integrals which enclose the real axis and the axis $z = -i\omega_m + \omega_1$, where $\omega_1$ is a real frequency. The result is

$$
\begin{aligned}
\Pi_1(i\omega_m) = \int \frac{d\omega_1}{4i\pi} \Big\{ &\left[ G_{\hat{a}}(\omega_1 + i0^+) - G_{\hat{a}}(\omega_1 - i0^+)\right]\\
\times\, & G_\eta(\omega_1 + i\omega_m) + \left[G_\eta(\omega_1 + i0^+) - G_\eta(\omega_1 - i0^+)\right]\\
\times\, & G_{\hat{a}}(\omega_1 - i\omega_m) \Big\} \tanh\left(\frac{\beta\omega_1}{2}\right).
\end{aligned}
\quad (22)
$$

Performing the analytical continuation $i\omega_m \to \omega + i0^+$ and substituting the unperturbed Green's functions Eq. (18), we find

$$
\begin{aligned}
\Pi_1(\omega) = \int \frac{d\omega_1}{4\pi v_F}\frac{\tanh(\beta\omega_1/2)}{\omega_1^2 + \Gamma^2}\\
\left[\frac{i\Gamma(\omega_1 + \omega)}{\omega_1 + \omega + i\Gamma} - \frac{\omega_1^2}{\omega_1 - \omega - i\Gamma}\right].
\end{aligned}
\quad (23)
$$

This integral is cutoff at $|\omega_1| \sim E_C$ where the decoupling of charge and spin excitations breaks down. The second term $\Pi_2$ is calculated along the same line with the result

$$\Pi_2(\omega) = \int \frac{d\omega_1}{2\pi v_F}\tanh\left(\frac{\beta\omega_1}{2}\right)\frac{\Gamma\omega_1}{\omega_1^2 + \Gamma^2}\frac{\Gamma - i\omega}{\omega_1^2 + (\Gamma - i\omega)^2}. \quad (24)$$

At zero temperature, the sum of the two integrals $\Pi_1$ and $\Pi_2$ is carried out to give the crossover function Eq. (7) from the main text.

---